\documentclass[aps,prd,twocolumn,superscriptaddress,showpacs]{revtex4-1}

\usepackage[dvips]{graphicx}
\usepackage{ulem}
\usepackage[usenames]{color}
\usepackage{amsmath}
\usepackage{epstopdf}

\usepackage{float}

\begin{document}


\title{Theoretical interpretation of $\Xi_c(2970)$}


\author{Ze Zhao}
\email{zhaoze@mail.itp.ac.cn}
\affiliation{CAS Key Laboratory of Theoretical Physics, Institute of Theoretical Physics, Chinese Academy of Sciences, Beijing 100190, China}


\begin{abstract}
The open charm strong decay widths and certain ratio of branching fractions of a charmed strange baryon $\Xi_c(2970)$ are calculated in a $^3P_0$ model. The results are compatible with the latest experimental data. The theoretical ratio of decay branching fractions $R = \mathcal{B} [ \Xi_{c}(2970)^+ \rightarrow \Xi_c(2645)^{0}\pi^{+} ] / \mathcal{B} [ \Xi_{c}(2970)^+ \rightarrow \Xi_c^{\prime0}\pi^{+}] \approx 1.0$. The spin-parity $J^P = 1/2^+$ and $3/2^+$ for different assignments are analyzed. From the results of our calculation, $\Xi_{c}(2970)$ can be interpreted as a $2S$-wave state with $J^P(s_l) = 1/2^+(0)$. The distinguishing between the 2S-wave $n_\rho$- and $n_\lambda$-excitation states and between states with $s_l = 0$ and $s_l = 1$ and between states with total spin $1/2$ and $3/2$($s_l = 1$) are also discussed.

\end{abstract}


\maketitle

\section{Introduction \label{sec:introduction}}
In the convention of quark model~\cite{pdg,klempt}, $\Xi_c$ baryon is consisted of with one $u$ or $d$, one strange and one charmed quark, which is also known as the charmed-strange baryon. The low lying charmed-strange baryons are confirmed with their $J^P$ numbers measured experimentally. However, with the increasing energy, more and more highly excited charmed strange baryons have been observed. In $\Xi_c$ sector, $\Xi_c(2970)$(was $\Xi_c(2980)$) was first observed by the Belle Collaboration in the $\Lambda_c^+K^-\pi^+$ channel~\cite{Chistov:0606051}, and then confirmed by the BABAR Collaboration in the $\Xi_c(2645)^0\pi^+$ decay channel~\cite{Aubert:0710.5763}. With its quantum numbers unmeasured, calculations and debates of $\Xi_c(2970)$ have been carried out. 

Recently, the Belle Collaboration reported the results from a study of the spin and parity of $\Xi_c(2970)^+$~\cite{Moon:2020gsg}. The angular distributions strongly favors $\Xi_c(2970)$ to be spin $J=1/2$. And they also measured the ratio of decay branching fractions, 

\begin{equation*}
\begin{split}
  R &= \frac{\mathcal{B} [ \Xi_{c}(2970)^+ \rightarrow \Xi_c(2645)^{0}\pi^{+} ] }{ \mathcal{B} [ \Xi_{c}(2970)^+ \rightarrow \Xi_c^{\prime0}\pi^{+}] }\\
    &= 1.67 \pm 0.29(\text{stat.})^{+0.15}_{-0.09}(\text{syst.}) \pm 0.25(\text{IS}) .
\end{split}
\end{equation*}

where the IS means the uncertainty due to possible isospin-symmetry-breaking effects. This R value favors the spin-parity $J^P = 1/2^+$ with the spin of the light-quark degrees of freedom $s_l = 0$.

In experiments, $J^P$ quantum numbers for most of the observed excited charmed baryons have not yet been measured so far. How to identify the observed baryons is an important topic in baryon spectroscopy. The spectroscopy of charmed baryons has been studied in many models. Ebert $\it{et~al}$  calculated the mass spectra of heavy baryons in the heavy-quark-light-quark picture in the QCD-motivated relativistic quark model and they suggested that $\Xi_c(2970)$ be assigned as the 2S excitation with $J^P = 1/2^+$~\cite{D Ebert:1105.0583}. Bing Chen $\it {et al}$ investigated the  $\Lambda_c$ and $\Xi_c$ in the heavy quark-light diquark picture and they also concluded that $\Xi_c(2970)$ to be assigned  to the first radial excitations with $J^P = 1/2^+$~\cite{chenbing2015}. For more spectral study, one can see literature~\cite{klempt} and references therein. 

Hadronic decays of $\Xi_c(2970)$ have been studied in a heavy hadron chiral perturbation theory~\cite{cheng}, where $\Xi_c(2970)$ is suggested a positive-parity excitation of $\Xi_c$.
It has also been studied in a chiral quark model~\cite{zhong}, in which $\Xi_c(2970)$ is suggested as one of the orbital excitations of $\Xi_c$. There is never certain conclusions about these excited baryons before their quantum numbers be measured experimentally. Since the report of the spin-parity of  $\Xi_c(2970)^+$, one of the remained questions is to ascertain its radial quantum number, that is to decide whether it is a 2S-wave or 1D-wave state. One of the purposes of this article is to solve this question theoretically by calculating the hadronic decay of $\Xi_c(2970)^+$ under different assignments, and the other is to predict the  properties of other $\Xi_c$ baryons that may carry similar masses or quantum numbers to be observed.

There are a lot of theoretical approaches to study the properties of hadrons. $^3P_0$ model is one of the phenomenological methods to calculate the OZI-allowed hadronic decays of hadrons. In addition to mesons, it is employed successfully to explain the hadronic decays of baryons~\cite{Capstick:2809 (1986),Roberts:171 (1992),Capstick:1994 (1993),Capstick:4507 (1994),Capstick:S241 (2000),Chong:094017 (2007)}. In this paper, we will study the hadronic decays of $\Xi_c(2970)^+$ in the framework of $^3P_0$ model.

This work is organized as follows. In Sec. II, we give a brief review of the $^3P_0$ model while in Sec. III we present our numerical results . In the last section, we give our conclusions and discussions.

\section{The framework \label{Sec: $^3P_0$ model}}
$^3P_0$ model was first proposed by Micu\cite{micu1969} and further developed by Yaouanc $\it{et~al~}$ ~\cite{yaouanc1,yaouanc2,yaouanc3}. It is also known as a Quark Pair Creation (QPC) model, in whose framework it assumes that a pair of quarks $q\bar{q}$ is created from the vacuum and thus with quantum numbers $J^{PC} = 0^{++}$($^{2S+1}L_J = ^3P_0$). It was first proposed to calculate the open strong decays of two-body mesons. Furthermore, the model has been subsequently employed and developed to study the OZI-allowed hadronic decays of three-body baryons by many authors not cited here.

In the model, the created quark-anti-quark $q\bar{q}$ then regroup with the quarks from the initial hadron A to form two daughter hadrons B and C. The interaction Hamiltonian for the creating process has the form~\cite{Geiger1994,Ackleh1996,Close2005}
\begin{eqnarray}
\mathcal{H}_{q \bar q} = \gamma \sum_f 2 m_f \int d^{\, 3} \bf{r} \bar{\psi}_{\it{f}} \psi_{\it{f}} ,
\end{eqnarray}
where $\psi_f$ is a Dirac quark field with flavor $f$. $m_f$ is the constituent quark mass. The strength of the quark pair creation is represented by the dimensionless parameter $\gamma$.

\begin{figure}
\begin{center}
\includegraphics[height=2.8cm,angle=0,width=6cm]{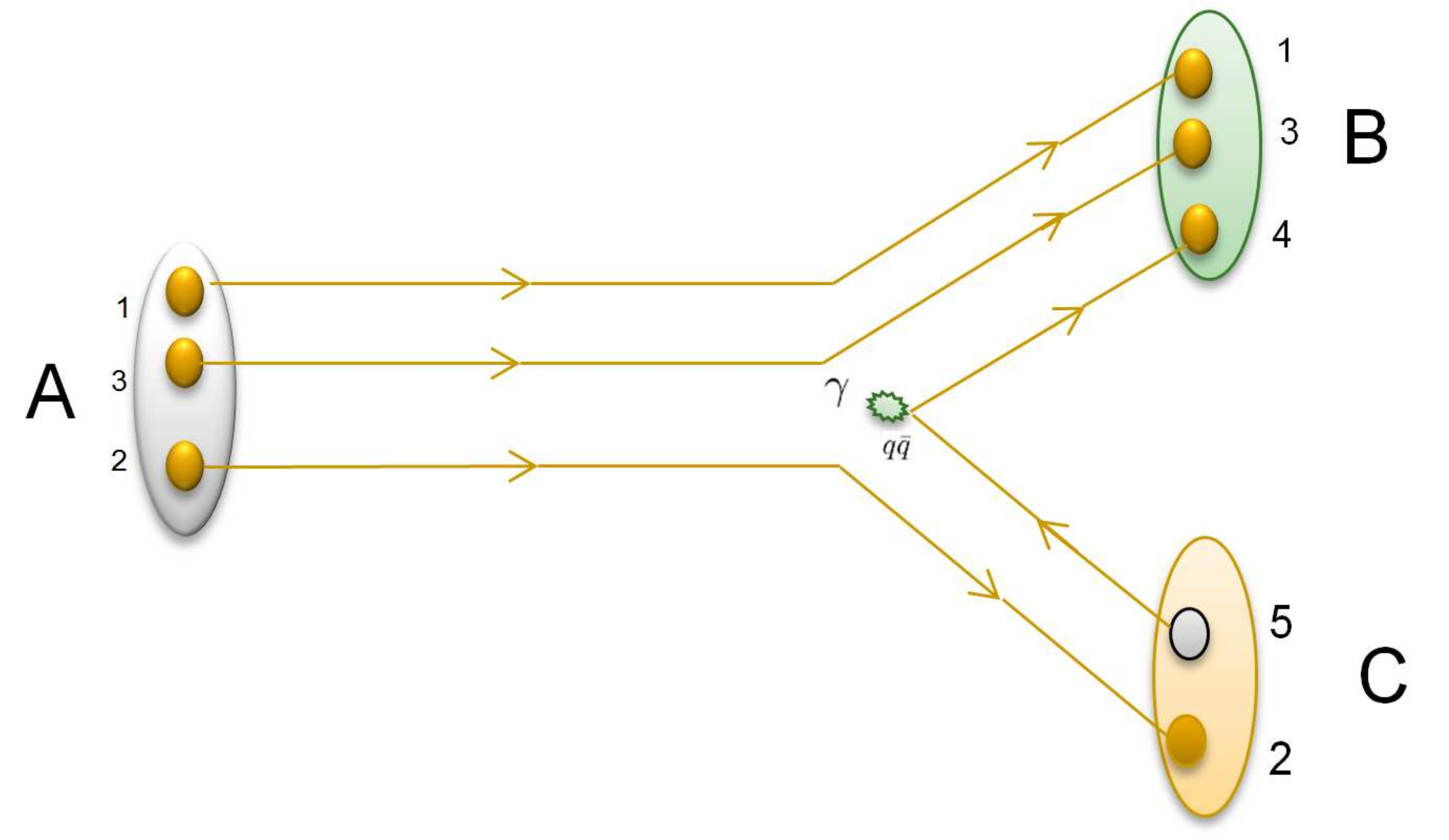}
\caption{Baryon decay process of $A\to B+C$ in the $^3P_0$ model.}
\label{3p0}
\end{center}
\end{figure}

For meson decays, the created quark regroup with the anti-quark of the initial meson, the created anti-quark regroup with the quark of the initial meson, and two mesons appear in the final states. For baryon decays, one quark of the initial baryon regroups with the created anti-quark to form a meson, and the rest two quarks regroup with the created quark to form a daughter baryon. The process of a baryon decay is shown in Fig.~\ref{3p0}.

In the $^3P_0$ model, the hadronic decay width $\Gamma$ of a process $A \to B + C$ is as follows~\cite{yaouanc3},
\begin{eqnarray}
\Gamma  = \pi ^2 \frac{|\vec{p}|}{m_A^2} \frac{1}{2J_A+1}\sum_{M_{J_A}M_{J_B}M_{J_C}} |{\mathcal{M}^{M_{J_A}M_{J_B}M_{J_C}}}|^2.
\end{eqnarray}
In the equation, $\vec{p}$ is the momentum of the daughter baryon in A's center of mass frame,
\begin{eqnarray}
 |\vec{p}|=\frac{{\sqrt {[m_A^2-(m_B-m_C )^2][m_A^2-(m_B+m_C)^2]}}}{{2m_A, }}
\end{eqnarray}
$m_A$ and $J_A$ are the mass and total angular momentum of the initial baryon A, respectively. $m_B$ and $m_C$ are the masses of the final hadrons. $\mathcal{M}^{M_{J_A}M_{J_B}M_{J_C}}$ is the helicity amplitude, which has the relation~\cite{Chong:094017 (2007)}
\begin{flalign}\label{heli_am}
 &\delta^3(\vec{p_B}+\vec{p_C}-\vec{p_A})\mathcal{M}^{M_{J_A } M_{J_B } M_{J_C }}\nonumber \\
 &=-2\gamma\sqrt {8E_A E_B E_C }  \sum_{M_{\rho_A}}\sum_{M_{L_A}}\sum_{M_{\rho_B}}\sum_{M_{L_B}} \sum_{M_{S_1},M_{S_3},M_{S_4},m}  \nonumber\\
 &\langle {J_{l_A} M_{J_{l_A} } S_3 M_{S_3 } }| {J_A M_{J_A } }\rangle \langle {L_{\rho_A} M_{L_{\rho_A} } L_{\lambda_A} M_{L_{\lambda_A} } }| {L_A M_{L_A } }\rangle \nonumber \\
 &\langle L_A M_{L_A } S_{12} M_{S_{12} }|J_{l_A} M_{J_{l_A} } \rangle \langle S_1 M_{S_1 } S_2 M_{S_2 }|S_{12} M_{S_{12} }\rangle \nonumber \\
 &\langle {J_{l_B} M_{J_{l_B} } S_3 M_{S_3 } }| {J_B M_{J_B } }\rangle \langle {L_{\rho_B} M_{L_{\rho_B} } L_{\lambda_B} M_{L_{\lambda_B} } }| {L_B M_{L_B } }\rangle \nonumber \\
 &\langle L_B M_{L_B } S_{14} M_{S_{14} }|J_{l_B} M_{J_{l_B} } \rangle \langle S_1 M_{S_1 } S_4 M_{S_4 }|S_{14} M_{S_{14} }\rangle \nonumber \\
 &\langle {1m;1 - m}|{00} \rangle \langle S_4 M_{S_4 } S_5 M_{S_5 }|1 -m \rangle \nonumber \\
 &\langle L_C M_{L_C } S_C M_{S_C}|J_C M_{J_C} \rangle \langle S_2 M_{S_2 } S_5 M_{S_5 }|S_C M_{S_C} \rangle \nonumber \\
&\times\langle\varphi _B^{1,4,3} \varphi _C^{2,5}|\varphi _A^{1,2,3}\varphi _0^{4,5} \rangle \times I_{M_{L_B } ,M_{L_C } }^{M_{L_A },m} (\vec{p}).
\end{flalign}
In the equation above, $\langle \varphi_B^{1,4,3} \varphi_C^{2,5}|\varphi_A^{1,2,3}\varphi_0^{4,5} \rangle$ is the flavor matrix, to calculate the flavor matrix element, equations from Ref.~\cite{yaouanc3} is employed,
\begin{flalign}
\langle \varphi_B^{1,4,3} \varphi_C^{2,5}|\varphi_A^{1,2,3}\varphi_0^{4,5} \rangle \nonumber \\
&=  \mathcal{F}^{(I_A;I_BI_C)}<I_Bi_BI_Ci_C|I_Ai_A>
\end{flalign}
and taking into account that $I_P=0$(isospin of the created quark pair) we have,
\begin{flalign}
\mathcal{F}^{(I_A;I_BI_C)} \nonumber \\
&= f \cdot (-1)^{I_{12}+I_C+I_A+I_3} \nonumber \\
&\times [\frac{1}{2}(2I_C  + 1)(2I_B  + 1)]^{1/2} \nonumber \\
&\times \begin{Bmatrix}
    {I_{12}} & {I_B} & {I_4}\\
    {I_C} & {I_3} & {I_A}\\ \end{Bmatrix}
\end{flalign}
where $f$ takes the value of $(\frac{2}{3})^{1/2}$ or $-(\frac{1}{3})^{1/2}$ according to the isospin $\frac{1}{2}$ or $0$ of the created quarks. $I_{A}$, $I_B$ and $I_M$ represent the isospins of the initial baryon, the final baryon and the final meson. $I_{12}$, $I_{3}$, $I_{4}$ are the isospins of relevant quarks, respectively. 

The core of the calculation lies in the space integral in Eq.(\ref{heli_am}),
\begin{flalign}
I_{M_{L_B } ,M_{L_C } }^{M_{L_A } ,m} (\vec{p})&= \int d \vec{p}_1 d \vec{p}_2 d \vec{p}_3 d \vec{p}_4 d \vec{p}_5 \nonumber \\
&\times\delta ^3 (\vec{p}_1 + \vec{p}_2 + \vec{p}_3 -\vec{p}_A)\delta ^3 (\vec{p}_4+ \vec{p}_5)\nonumber \\
&\times \delta ^3 (\vec{p}_1 + \vec{p}_4 + \vec{p}_3 -\vec{p}_B )\delta ^3 (\vec{p}_2 + \vec{p}_5 -\vec{p}_C) \nonumber \\
& \times\Psi _{B}^* (\vec{p}_1, \vec{p}_4,\vec{p}_3)\Psi _{C}^* (\vec{p}_2 ,\vec{p}_5) \nonumber \\
& \times \Psi _{A} (\vec{p}_1 ,\vec{p}_2 ,\vec{p}_3)y _{1m}\left(\frac{\vec{p_4}-\vec{p}_5}{2}\right).
\end{flalign}

Simple harmonic oscillator (SHO) wave functions are employed to model the baryon wave functions\cite{Capstick:2809 (1986),Capstick:1994 (1993),Capstick:4507 (1994)}
\begin{flalign}
\Psi_{A}(\vec{p}_{A})&=N\Psi_{n_{\rho_A} L_{\rho_A} M_{L_{\rho_A}}}(\vec{p}_{\rho_A}) \Psi_{n_{\lambda_A} L_{\lambda_A} M_{L_{\lambda_A}}}(\vec{p}_{\lambda_A}),
\end{flalign}
\begin{flalign}
\Psi_{B}(\vec{p}_{B})&=N\Psi_{n_{\rho_B} L_{\rho_B} M_{L_{\rho_B}}}(\vec{p}_{\rho_B}) \Psi_{n_{\lambda_B} L_{\lambda_B} M_{L_{\lambda_B}}}(\vec{p}_{\lambda_B}),
\end{flalign}
where $N$ represents a normalization coefficient of the total wave function and
\begin{flalign}
\Psi_{nLM_L}(\vec{p})&=\frac{(-1)^n(-i)^L}{\beta^{3/2}}\sqrt{\frac{2n!}{\Gamma(n+L+\frac{3}{2})}}\big(\frac{\vec{p}}{\beta}\big)^L \exp(-\frac{\vec{p}^2}{2\beta^2}) \nonumber\\
&\times L_n^{L+1/2}\big(\frac{\vec{p}^2}{\beta^2}\big)Y_{LM_L}(\Omega_p).
\end{flalign}
$L_n^{L+1/2}\big(\frac{\vec{p}^2}{\beta^2}\big)$ denotes the Laguerre polynomial function, $Y_{LM_L}(\Omega_p)$ is a spherical harmonic function. The relation between the solid harmonica polynomial $y _{LM}(\vec{p})$ and $Y_{LM_L}(\Omega_{\vec{p}})$ is $y _{LM}(\vec{p})=|\vec{p}|^L Y_{LM_L}(\Omega_p)$.

\section{Numerical results \label{Sec: Numerical results}}
\subsection{Notations of baryons and relevant parameters}

For the hadronic decays of $\Xi_c(2970)$, the quantum numbers of the initial state baryons and final state baryons are presented in Table I and Table II, respectively. Here $n_\rho$, $L_\rho$ and $S_\rho$ denote the nodal, the orbital and the spin of the two light quarks, while $n_\lambda$, $L_\lambda$ and $S_\lambda$ correspond to the nodal, the orbital and the spin between the heavy quark and the light quark system. $J_l$ is the total angular momentum of  $S_\rho$ and the total orbital angular momentum $L$. Finally, $J$ equals to the total spin of the baryon. Notations for the excited D-wave $\Xi_c$ baryons are the same as those in Ref.~\cite{Chong:094017 (2007)}. For the first radially excited $\Xi_c$, there are two kinds of excitations, ($n_\rho$, $n_\lambda$) = (1, 0) and (0, 1), which we call them as ``$n_\rho$-excitation" and ``$n_\lambda$-excitation", respectively. We use $\tilde{\Xi}_{cJ_l}^{(\prime,*)}$ and $\acute{\Xi}_{cJ_l}^{(\prime,*)}$ to represent them.

\begin{table}[htbp]
	\caption{Quantum numbers of 2S- and 1D-wave excitations}
	\begin{tabular}{p{0.0cm} p{2.5cm}*{8}{p{0.5cm}}}
		\hline\hline
		& Assignments                                         & $J$                         & $J_l$ & $n_\rho$ & $L_\rho$ & $n_\lambda$ & $L_\lambda$ & $L$  & $S_\rho$ \\
		
		\hline
		\label{40}   &$\tilde{\Xi}_{c0}(\frac{1}{2}^+)$                   & $\frac{1}{2}$   &0    &  1 & 0  & 0 &  0 &  1 &  0  \\
		\label{38}   &$\tilde{\Xi}_{c1}^{'}(\frac{1}{2}^+)$             & $\frac{1}{2}$   &1    &  1 & 0  & 0 &  0 &  1 &  1  \\		
		\label{39}   &$\tilde{\Xi}_{c1}^{*}(\frac{3}{2}^+)$            & $\frac{3}{2}$   &1    &  1 & 0  & 0 &  0 &  1 &  1  \\
		
		\label{37}   &$\acute{\Xi}_{c0}(\frac{1}{2}^+)$                   & $\frac{1}{2}$   &0    &  0 & 0  & 1 &  0 &  1 &  0  \\
		\label{35}   &$\acute{\Xi}_{c1}^{'}(\frac{1}{2}^+)$             & $\frac{1}{2}$   &1    &  0 & 0  & 1 &  0 &  1 &  1  \\
		\label{36}   &$\acute{\Xi}_{c1}^{*}(\frac{3}{2}^+)$            &$\frac{3}{2}$    &1    &  0 & 0  & 1 &  0 &  1 &  1  \\

		\label{  }   &                                                    &                 &     &    &    &   &    &    &     \\

		\label{01}   &$\Xi_{c1}^{' }(\frac{1}{2}^+,\frac{3}{2}^+)$        & $\frac{1}{2}$,$\frac{3}{2}$ &  1    &  0       &  0       &  0       &   2         &  2   &  1       \\
		\label{03}   &$\Xi_{c2}^{' }(\frac{3}{2}^+,\frac{5}{2}^+)$        & $\frac{3}{2}$,$\frac{5}{2}$ &  2    &  0       &  0       &  0       &   2         &  2   &  1       \\
		\label{05}   &$\Xi_{c3}^{' }(\frac{5}{2}^+,\frac{7}{2}^+)$        & $\frac{3}{2}$,$\frac{5}{2}$ &  3    &  0       &  0       &  0       &   2         &  2   &  1       \\
		\label{07}   &$\Xi_{c2}^{  }(\frac{3}{2}^+,\frac{5}{2}^+)$        & $\frac{3}{2}$,$\frac{5}{2}$ &  2    &  0       &  0       &  0       &   2         &  2   &  0       \\
		\label{09}   &$\hat\Xi_{c1}^{' }(\frac{1}{2}^+,\frac{3}{2}^+)$    & $\frac{1}{2}$,$\frac{3}{2}$ &  1    &  0       &  2       &  0       &   0         &  2   &  1       \\
		\label{11}   &$\hat\Xi_{c2}^{' }(\frac{3}{2}^+,\frac{5}{2}^+)$    & $\frac{3}{2}$,$\frac{5}{2}$ &  2    &  0       &  2       &  0       &   0         &  2   &  1       \\
		\label{13}   &$\hat\Xi_{c3}^{' }(\frac{5}{2}^+,\frac{7}{2}^+)$    & $\frac{3}{2}$,$\frac{5}{2}$ &  3    &  0       &  2       &  0       &   0         &  2   &  1       \\
		\label{15}   &$\hat\Xi_{c2}^{ }(\frac{3}{2}^+,\frac{5}{2}^+)$     & $\frac{3}{2}$,$\frac{5}{2}$ &  2    &  0       &  2       &  0       &   0         &  2   &  0       \\
		\label{17}   &$\check\Xi_{c0}^{'0}(\frac{1}{2}^+)$                & $\frac{1}{2}$               &  0    &  0       &  1       &  0       &   1         &  0   &  0       \\
		\label{18}   &$\check\Xi_{c1}^{'1}(\frac{1}{2}^+,\frac{3}{2}^+)$  & $\frac{1}{2}$,$\frac{3}{2}$ &  1    &  0       &  1       &  0       &   1         &  1   &  0       \\
		\label{20}   &$\check\Xi_{c2}^{'2}(\frac{3}{2}^+,\frac{5}{2}^+)$  & $\frac{3}{2}$,$\frac{5}{2}$ &  2    &  0       &  1       &  0       &   1         &  2   &  0       \\
		\label{22}   &$\check\Xi_{c1}^{\ 0}(\frac{1}{2}^+,\frac{3}{2}^+)$ & $\frac{1}{2}$,$\frac{3}{2}$ &  1    &  0       &  1       &  0       &   1         &  0   &  1       \\
		\label{24}   &$\check\Xi_{c0}^{\ 1}(\frac{1}{2}^+)$               & $\frac{1}{2}$               &  0    &  0       &  1       &  0       &   1         &  1   &  1       \\
		\label{25}   &$\check\Xi_{c1}^{\ 1}(\frac{1}{2}^+,\frac{3}{2}^+)$ & $\frac{1}{2}$,$\frac{3}{2}$ &  1    &  0       &  1       &  0       &   1         &  1   &  1       \\
		\label{27}   &$\check\Xi_{c2}^{\ 1}(\frac{3}{2}^+,\frac{5}{2}^+)$ & $\frac{3}{2}$,$\frac{5}{2}$ &  2    &  0       &  1       &  0       &   1         &  1   &  1       \\
		\label{29}   &$\check\Xi_{c1}^{\ 2}(\frac{1}{2}^+,\frac{3}{2}^+)$ & $\frac{1}{2}$,$\frac{3}{2}$ &  1    &  0       &  1       &  0       &   1         &  2   &  1       \\
		\label{31}   &$\check\Xi_{c2}^{\ 2}(\frac{3}{2}^+,\frac{5}{2}^+)$ & $\frac{3}{2}$,$\frac{5}{2}$ &  2    &  0       &  1       &  0       &   1         &  2   &  1       \\
		\label{33}   &$\check\Xi_{c3}^{\ 2}(\frac{5}{2}^+,\frac{7}{2}^+)$ & $\frac{5}{2}$,$\frac{7}{2}$ &  3    &  0       &  1       &  0       &   1         &  2   &  1       \\

		\hline\hline
	\end{tabular}
	\label{table1}
\end{table}

\begin{table}[t]
\caption{Quantum numbers of baryons in the final states }
\begin{tabular}{p{0.0cm} p{2.4cm}*{6} {p{0.8cm}}}
   \hline\hline
   &$            $       & $J$           & $J_l$ & $L_\rho$ & $L_\lambda$ & $L$  & $S_\rho$ \\
   \hline
   &$\Xi_c^{0(+)}$       & $\frac{1}{2}$ &  0    &  0       &   0         &  0   &  0  \\
   &$\Xi_c^{'0(+)}$      & $\frac{1}{2}$ &  1    &  0       &   0         &  0   &  1  \\
   &$\Xi_c(2645)^{0(+)}$      & $\frac{3}{2}$ &  1    &  0       &   0         &  0   &  1  \\
   &$\Sigma_c(2455)^{+(++)}$ & $\frac{1}{2}$ &  1    &  0       &   0         &  0   &  1  \\
   &$\Lambda_c^+$        & $\frac{1}{2}$ &  0    &  0       &   0         &  0   &  0  \\

   \hline\hline
\end{tabular}
\label{QNFI}
\end{table}

\begin{table}[t]
	\caption{Masses of involved mesons and baryons in the decays~\cite{pdg}}
	\begin{tabular}{p{0.0cm} p{2.0cm}p{2.0cm}|p{2.0cm}p{2.0cm}}
		\hline\hline
		&State             &mass (MeV) & State               &mass (MeV)\\
		\hline
		&$\pi^{\pm}      $ &139.570    &$\Xi_c(2645)^0    $  &2646.38  \\
		&$\pi^{0}        $ &134.977    &$\Xi_c(2645)^+    $  &2645.56  \\
		&$K^{\pm}        $ &493.677    &$\Xi_c(2790)^0    $  &2794.1  \\
		&$K^{0}          $ &497.611    &$\Xi_c(2790)^+    $  &2793.4  \\
		&$\Xi_c^0        $ &2470.90    &$\Xi_c(2815)^0    $  &2820.25  \\
		&$\Xi_c^+        $ &2467.94    &$\Xi_c(2815)^+    $  &816.74  \\
		&$\Xi_c^{'0}     $ &2579.2     &$\Sigma_c(2455)^{+}$ &2452.9   \\
		&$\Xi_c^{'+}     $ &2578.4     &$\Sigma_c(2455)^{++}$&2453.97   \\
		&$\Lambda_c^{+} $ &2286.46    &                     &        \\
		
		\hline\hline
	\end{tabular}
	\label{masses}
\end{table}

As reported by the Belle Collaboration\cite{Moon:2020gsg}, the $J^P$ quantum numbers of $\Xi_c(2970)$ favor $J^P(s_l) = 1/2^+(0)$. From Table~\ref{QNFI} we see that the possible assignments for $\Xi_c(2970)$ are the 2S-wave $\tilde{\Xi}_{c0}(\frac{1}{2}^+)$ and $\acute{\Xi}_{c0}(\frac{1}{2}^+)$, the 1D-wave $\check\Xi_{c0}^{'0}(\frac{1}{2}^+)$ and $\check\Xi_{c1}^{'1}(\frac{1}{2}^+)$.

Masses of relevant mesons and baryons involved in the calculation are listed in Table \ref{masses}\cite{pdg}.

Parameters $\gamma$ and $\beta$ are taken as those in Refs.~\cite{Chong:094017 (2007),Blundell:3700 (1996)}. $\gamma=13.4$. $\beta$ is chosen as $476$ MeV for meson $\pi$ and $K$. For baryons, a universal value $\beta=600$ MeV is employed. The mass of $\Xi_c(2970)^0$ is taken as $2966.34$ MeV.

\subsection{Decays of $\Xi_c(2970)^+$ as 2S- and 1D-wave state}
In general, $u\bar u$, $d\bar d$ and $s\bar s$ could be created from the vacuum. However, there exists no experimental signal for the decay mode with a $s\bar s$ creation. On the other hand, according to the decaying  threshold the decay mode with a $s\bar s$ creation does not open for $\Xi_c(2970)^0$. Thus, we consider that the OZI-allowed channels are all assumed from the $u\bar u$ and $d\bar d$ pairs creation. Possible decay modes and corresponding hadronic decay widths of $\Xi_c(2970)^+$ as 2S- and 1D-wave state baryons in different assignments are computed and presented in Table~\ref{results}, along with the decays of a few other candidate-like states are also calculated and listed. The vanish modes the table indicate forbidden channels due to conservation of some quantum numbers.

\begin{table*}[t]
\begin{center}
	\caption{Decay widths (MeV) of $\Xi_c(2970)^+$ as different states. .}
	\begin{tabular}{p{2.3cm}p{1.5cm}p{1.5cm}p{1.5cm}p{1.5cm}p{1.5cm}p{1.5cm}p{1.5cm}p{1.5cm}}
		\hline \hline
States                 &$\check\Xi_{c0}^{'0+}(\frac{1}{2}^+)$ &$\check\Xi_{c1}^{'1+}(\frac{1}{2}^+)$ &$\tilde\Xi_c^{+}(\frac{1}{2}^+)$ &$\tilde\Xi_c^{\prime+}(\frac{1}{2}^+)$ &$\tilde\Xi_c^{*+}(\frac{3}{2}^+)$   &$\acute{\Xi}_c^{+}(\frac{1}{2}^+)$ &$\acute{\Xi}_c^{\prime+}(\frac{1}{2}^+)$ &$\acute{\Xi}_c^{*+}(\frac{3}{2}^+)$ \\
		\hline
$\Xi_c^0\pi^+    $                 &0       &0   &0     &71.7                &71.7                &0                 &8.0                 &8.0                 \\ 
$\Xi_c^{'0}\pi^+ $               &22.3   &0   &33.4  &44.6                &11.1                &3.7               &5.0                 &1.2                 \\ 
$\Xi_c(2645)^{0}\pi^+$       &22.1   &0   &33.2  &11.1                &27.7                &3.7               &1.2                 &3.1                 \\ 
$\Xi_c(2790)^{0}\pi^+$       &0       &0   &0     &4.9                 &$6.0\times 10^{-5}$ &0                 &7.0                 &$1.3\times 10^{-3}$ \\ 
$\Xi_c(2815)^{0}\pi^+$       &0       &0   &0     &$1.6\times 10^{-6}$ &1.8                 &0                 &$3.3\times 10^{-5}$ &2.7                 \\ 
$\Xi_c^+\pi^0    $                &0        &0   &0     &73.3                &73.3                &0                 &8.1                 &8.1                 \\ 
$\Xi_c^{'+}\pi^0 $              &22.7    &0   &34.0  &45.4                &11.3                &3.8               &5.0                 &1.3                 \\ 
$\Xi_c(2645)^{+}\pi^0$      &22.8    &0   &34.2  &11.4                &28.5                &3.8               &1.3                 &3.2                 \\ 
$\Xi_c(2790)^{+}\pi^0$      &0        &0   &0     &5.3                 &$9.1\times 10^{-5}$ &0                 &7.6                 &$1.9\times 10^{-3}$ \\ 
$\Xi_c(2815)^{+}\pi^0$      &0        &0   &0     &$1.2\times 10^{-5}$ &2.8                 &0                 &$2.5\times 10^{-4}$ &4.0                 \\ 
$\Sigma_c(2455)^{+}K^0$ &1.5     &0   &2.2   &3.0                 &0.7                 &0.2               &0.3                 &0.1                 \\ 
$\Sigma_c(2455)^{++}K^-$&1.9    &0   &2.8   &3.8                 &0.9                 &0.3               &0.4                 &0.1                 \\ 
$\Lambda_c^{+}K^0$        &0       &0   &0     &81.6                &81.6                &0                 &9.1                 &9.1                 \\ 
Total                                  &93.3     &0   &140.0     &356.1                &311.4                &15.5                 &53.0                 &40.9                 \\ 
		\hline\hline
	\end{tabular}
	\label{results}
\end{center}
\end{table*}

$\Xi_c(2970)$ was first observed by the Belle Collaboration in the $\Lambda_c^+K^-\pi^+$ channel with $\Gamma = 43.5 \pm 7.5 \pm 7.0 MeV$ and confirmed by the BABAR Collaboration in the intermediate resonant mode $\Sigma_c(2455)^+K^-$. In Ref.\cite{Yelton2016}, some new measurements were reported. The branching fraction ratio of $B(\Xi_{c}(2970)^{+} \to \Xi_c^{'0} \pi^{+})/B(\Xi_{c}(2815)^{+} \to \Xi_{c}(2645)^{0}\pi^{+}, \Xi_{c}(2645)^{0} \to \Xi_c^+ \pi^-) \approx 75\% $ indicates that $\Xi_c(2970)$ decays significantly into $\Xi_c^\prime \pi^+$. The decay of $\Xi_c(2970)$ into $\Lambda_c K$ or $\Xi_c \pi$ channel has never been observed in experiments. The latest experimental data shows that the width of $\Xi_c(2970)^+$ is $\Gamma = 20.9^{+2.4}_{-3.5} MeV$

From Table \ref{results} we can see that for the four states with $s_l(S_\rho) = 0$, the D-wave $\check\Xi_{c0}^{'0+}(\frac{1}{2}^+)$ and $\check\Xi_{c1}^{'1+}(\frac{1}{2}^+)$, the 2S-wave $\tilde\Xi_c^{+}(\frac{1}{2}^+)$ and $\acute{\Xi}_c^{+}(\frac{1}{2}^+)$ have very different decay behaviors. The $\check\Xi_{c1}^{'1+}(\frac{1}{2}^+)$ does not decay to any of the channels. This may indicate that this state may never exist or because of the model itself or the spin coupling scheme in the baryons, however,  here we do not discuss it in detail.  The state $\check\Xi_{c0}^{'0+}(\frac{1}{2}^+)$ has the ratio R $\approx$ 1, which is near the lower bound of experimental value. On the other hand, it also has a total decay width of 93.3, which is about five times as the experimental measurement. 

For the 2S-wave states, the $n_\rho$-excited state $\tilde\Xi_c^{+}(\frac{1}{2}^+)$ also has the ratio R $\approx$ 1 and has a even more larger total decay width, which is about seven times as the experimental measurement. On the other hand, the $n_\lambda$-excited state $\acute{\Xi}_c^{+}(\frac{1}{2}^+)$ has a relatively smaller total decay width $15.5 MeV$, which is near the lower threshold of the experimental data. The R value is also $\approx$ 1.

On the other hand, as mentioned in the experimental article\cite{Moon:2020gsg}, heavy-quark spin symmetry(HQSS) predicts $R = 1.06 (0.26)$ for a $1/2^+$ state with the spin of the light-quark degrees of freedom $s_l = 0(1)$ as calculated in Ref~\cite{cheng}. The R value in this work are in consistent with their result, which is R $\approx$ 1.

As we all know, the results of phenomenological models depend heavily on the parameters. Ref.\cite{Ye:2017dra} calculated the decay widths of $\Xi_c$ baryons by using different groups of parameters $\beta$. The results in their Table IV shows how the decay width and relative branching fractions vary according the different sets of parameters.

Further more, we also calculated the hadronic decays of the 2S-wave $\Xi_c$ baryons with quantum numbers $s_l = 1$, the $n_\rho$-excitations $\tilde\Xi_c^{\prime+}(\frac{1}{2}^+)$ and $\tilde\Xi_c^{*+}(\frac{3}{2}^+)$, and the $n_\lambda$-excitations $\acute{\Xi}_c^{\prime+}(\frac{1}{2}^+)$ and $\acute{\Xi}_c^{*+}(\frac{3}{2}^+)$. From Table~\ref{results} we can see that the decay channels $\Xi_c\pi$ and $\Lambda_cK$ are open. Both these two channels have relatively larger decay widths than other channels. It can be taken as the distinguishing between state with $s_l = 0$ and $s_l = 1$. It is obviously that the total decay widths of $\tilde\Xi_c^{\prime+}(\frac{1}{2}^+)$ and $\tilde\Xi_c^{*+}(\frac{3}{2}^+)$ are much more larger than that of the $\acute{\Xi}_c^{\prime+}(\frac{1}{2}^+)$ and $\acute{\Xi}_c^{*+}(\frac{3}{2}^+)$, that is, the $n_\rho$-excited states are much more broader than the $n_\lambda$-excited states. The total decay widths will be the distinguishing between them. Finally, the branching ratios
$$\frac{\tilde\Xi_c^{\prime+}(\frac{1}{2}^+) \to \Xi_c(2645)^{0}\pi^+}{\tilde\Xi_c^{\prime+}(\frac{1}{2}^+) \to \Xi_c^{'0}\pi^+} = \frac{11.1}{44.6} \approx 0.25 < 0.5, $$
$$\frac{\tilde\Xi_c^{*+}(\frac{3}{2}^+) \to \Xi_c(2645)^{0}\pi^+}{\tilde\Xi_c^{*+}(\frac{3}{2}^+) \to \Xi_c^{'0}\pi^+} = \frac{27.7}{11.1} \approx 2.5 > 2 .$$
This may be of help to distinguish the 2S-wave states with total spin $\frac{1}{2}$ and $\frac{3}{2}$.

\section{Conclusions and discussions\label{Sec: summary}}

In this work, the hadronic decays of $\Xi_c(2970)^+$ are studied in a $^3P_0$ model. We calculate the decay widths and some ratios of branching fractions of $\Xi_c(2970)^+$ as possible assignments related to recent Belle experiment. In comparison with experiments, we make an identification of this charmed-strange baryon. Our theoretical predictions are consistent with experiments.

$\Xi_c(2970)$ was first observed by the Belle Collaboration in the $\Lambda_c^+K^-\pi^+$ channel, the latest experimental data shows that its decay width is of about 20 MeV. Our results of branching fraction ratios are in consistent with the results that considered the HQSS effects. Our conclusion of $\Xi_c(2970)^+$ is that it is a 2S-wave $n_\lambda$-excitation state $\acute{\Xi}_c^{+}(\frac{1}{2}^+)$, whose $J^P(s_l)$ is $1/2^+(0)$. It is in compatible with the latest experimental measurement. 

The 1D-wave $\check\Xi_{c0}^{'0+}(\frac{1}{2}^+)$ and $\check\Xi_{c1}^{'1+}(\frac{1}{2}^+)$ assignments are not suitable for $\Xi_c(2970)$ ether for over large total decay width or for no decays at all. The 2S-wave $\tilde\Xi_c^{+}(\frac{1}{2}^+)$ assignment is not suitable for $\Xi_c(2970)$ neither. The large difference of total decay widths may lead to different inner pictures of the $n_\rho$- and $n_\lambda$-excitations. However, the dynamics between this two pictures still require deeper investigations to reveal it.

We also calculate the hadronic decays of the 2S-wave $\Xi_c$ baryons with quantum numbers $s_l = 1$, the $n_\rho$-excitations $\tilde\Xi_c^{\prime+}(\frac{1}{2}^+)$ and $\tilde\Xi_c^{*+}(\frac{3}{2}^+)$, and the $n_\lambda$-excitations $\acute{\Xi}_c^{\prime+}(\frac{1}{2}^+)$ and $\acute{\Xi}_c^{*+}(\frac{3}{2}^+)$. The particular difference is that the decay channels $\Xi_c\pi$ and $\Lambda_cK$ are open, compared to the fact that the decay of $\Xi_c(2970)$ into $\Lambda_c K$ or $\Xi_c \pi$ channel has never been observed in experiments. It can be taken as the distinguishing between state with $s_l = 0$ and $s_l = 1$. Another feature we can see is that the $n_\rho$-excitation states $\tilde\Xi_c^{\prime+}(\frac{1}{2}^+)$ and $\tilde\Xi_c^{*+}(\frac{3}{2}^+)$ have much larger total decay widths than that of the $n_\lambda$-excitation states $\acute{\Xi}_c^{\prime+}(\frac{1}{2}^+)$ and $\acute{\Xi}_c^{*+}(\frac{3}{2}^+)$. This can also be a distinguishing between them. Finally, the branching fraction ratio $\frac{\Xi_c^{\prime+}(\frac{1}{2}^+) \to \Xi_c(2645)^{0}\pi^+}{\Xi_c^{\prime+}(\frac{1}{2}^+) \to \Xi_c^{'0}\pi^+} $ and $\frac{\Xi_c^{*+}(\frac{3}{2}^+) \to \Xi_c(2645)^{0}\pi^+}{\Xi_c^{*+}(\frac{3}{2}^+) \to \Xi_c^{'0}\pi^+} $ can be a dominant factor to distinguish the 2S-wave $\Xi_c^{\prime}(\frac{1}{2}^+)$ and $\Xi_c^{*}(\frac{3}{2}^+)$ with $s_l = 1$.

\begin{acknowledgments}
Ze Zhao thanks Prof. Ailin Zhang and Bing-Song Zou for helpful discussions and insightful suggestions. This work is supported by National Natural Science Foundation of China(NSFC) under Grant No. 11847225.
\end{acknowledgments}

\end{document}